# 後量子密碼學匿名憑證 PQCWC：

## Post-Quantum Cryptography Winternitz-Chen

Abel C. H. Chen

中華電信研究院 資通安全研究所 高級研究員

chchen.scholar@gmail.com

**摘要**

由於量子計算技術成熟將威脅到現行主流非對稱式密碼學方法(包含 RSA 密碼學和橢圓曲線密碼學)的安全性，所以美國國家標準暨技術研究院(National Institute of Standards and Technology, NIST)在 2024 年 8 月訂定了 3 個後量子密碼學演算法最終版標準文件，並且該些後量子密碼學演算法主要建構在晶格基礎(Lattice-base)和雜湊基礎(Hash-based)。有鑑於此，本研究提出 Post-Quantum Cryptography Winternitz-Chen (PQCWC)的匿名憑證方案，旨在探索後量子密碼學的基礎上設計匿名憑證方案以期未來可以應用在隱私保護場域。本研究設計的匿名憑證方案主要建構在 Winternitz 簽章方案的基礎，可以避免在憑證中曝露原本的公鑰。本研究亦把 PQCWC 匿名憑證方案結合蝴蝶金鑰擴展(Butterfly Key Expansion, BKE)機制，提出世界上第一個基於雜湊的蝴蝶金鑰擴展機制，達到對註冊中心(Registration Authority, RA)和憑證中心(Certificate Authority, CA)也能匿名，充份保護隱私。在實驗環境中，本研究比較各種不同的雜湊演算法，包含安全雜湊演算法 1(Secure Hash Algorithm-1, SHA-1)、SHA-2 系列、SHA-3 系列、以及 BLAKE 系列，證明提出的匿名憑證方案可以在不增加金鑰長度、簽章長度、產製金鑰時間、產製簽章時間、驗證簽章時間的情況下達到匿名性。

**關鍵字**：後量子密碼學、匿名憑證、基於雜湊密碼學、Winternitz 簽章方案。

## 一、前言

量子計算的技術已經日益成熟，搭配合適的量子演算法，計算效率將有機會達到指數級加速[1]。其中，知名演算法包含有 Shor 量子演算法[2]，可以通過量子傅利葉變換來快速找到隱藏子群及其週期[3]，將可以快速破解質因數分解問題和離散對數問題[4]。然而，RSA 密碼學和橢圓曲線密碼學等現行主流的非對稱式密碼學，主要建構在質因數分解問題和離散對數問題的基礎上來提供安全性，但這些安全性將可能被量子計算快速攻破[5]。因此，基於質因數分解問題和離散對數問題的非對稱式密碼學已不再安全[6]。有鑑於此，美國國家標準暨技術研究院(National Institute of Standards and Technology, NIST)近幾年來致力於訂定後量子密碼學(Post-Quantum Cryptography, PQC)標準[7]，並對全世界公開徵求收集基於晶格(lattice-based)密碼學[8]、基於雜湊(hash-based)密碼學[9]、基於編碼(code-based)密碼學[10]、基於多變量(multivariate-based)密碼學[11]等不同的演算法，並且在 2024 年 8 月確立了無狀態雜湊基礎數位簽章演算法(Stateless hash-based Digital Signature Algorithm, SLH-DSA)[12]、模晶格基礎數位簽章演算法(Module-Lattice-based Digital Signature Algorithm, ML-DSA)[13]、模晶格基礎金鑰封裝演算法(Module-Lattice-based Key Encapsulation Mechanism, ML-KEM)[14]等後量子

密碼學標準演算法。並且陸續有一些後量子密碼學的應用被發展出來，包含車聯網[15]、物聯網[16]、區塊鏈[17]等。其中，無狀態雜湊基礎數位簽章演算法是屬於基於雜湊密碼學的一種演算法，並且由於雜湊具有不可逆的特性，所以基於雜湊密碼學主要應用在數位簽章，而不適用於金鑰封裝。

目前的 X.509 憑證[18]和車聯網安全憑證管理系統憑證[19]等憑證格式都已經有標準及其規範，所以在更新為後量子密碼學演算法時主要修改憑證格式中的演算法物件識別碼(Object Identifier, OID)，並且在公鑰(public key)欄位放置其對應的待簽者公鑰，並且由簽發者(Issuer)對待簽署資料(to-be-signed-data)產製簽章資訊和放置在簽章(signature)欄位。然而，由於憑證中放置公鑰，在後續應用上將可能被攻擊者追蹤同一張憑證公鑰對應的使用者，並且分析其對應的同一把私鑰所簽發的訊息，而且分析出該使用者簽發的全部訊息，將導致有曝露使用者隱私的疑慮。有鑑於此，IEEE 1609.2.1 標準[20]中設計了蝴蝶金鑰擴展(Butterfly Key Expansion, BKE)機制[21]，通過蝴蝶金鑰擴展機制可以保護使用者原始的公鑰(在此機制中稱為毛蟲公鑰)，並且通過金鑰擴展的方式分別擴展成繭公鑰和蝴蝶公鑰，並且可以避免從蝴蝶公鑰反推出毛蟲公鑰，以保護使用者隱私。然而，在 IEEE 1609.2.1 標準中制定的蝴蝶金鑰擴展機制主要建構在橢圓曲線密碼學的基礎上，並且在橢圓曲線特性上做到金鑰擴展，而這種方式將可能被量子計算破解，所以無法提供量子安全(Quantum-Safe)等級。

有鑑於隱私保護和量子安全的需求，本研究在基於雜湊密碼學基礎上提出後量子密碼學匿名憑證方案，命名為 Post-Quantum Cryptography Winternitz-Chen (PQCWC)。本研究方法主要基於 Winternitz 一次性簽章方法[22]設計出匿名憑證方案，並且搭配蝴蝶金鑰擴展機制來提供隱私保護功能，在基於雜湊密碼學的特性上達到量子安全。本研究的貢獻條列如下：

- 本研究提出 Post-Quantum Cryptography Winternitz-Chen (PQCWC)演算法，可以在基於雜湊密碼學的特性提供匿名憑證方案，達到量子安全。
- 本研究全球首創的基於雜湊蝴蝶金鑰擴展機制(Hash-based Butterfly Key Expansion, HBKE)，可以做到對系統內的每個設備(包含註冊中心(Registration Authority, RA)和憑證中心(Certificate Authority, CA))都匿名，充份保護隱私。
- 本研究比較各種不同的雜湊演算法，包含安全雜湊演算法 1(Secure Hash Algorithm-1, SHA-1)、SHA-2 系列、SHA-3 系列、以及 BLAKE 系列，證明提出的匿名憑證方案可以在不增加金鑰長度、簽章長度、產製金鑰時間、產製簽章時間、驗證簽章時間的情況下達到匿名性。

本文總共分為六個章節。第二節將對現有方法進行文獻探討，並且介紹 Winternitz 一次性簽章方法、已知憑證請求和回應流程、IEEE 1609.2.1 蝴蝶金鑰擴展機制。第三節提出本研究設計的後量子密碼學匿名憑證方案，Post-Quantum Cryptography Winternitz-Chen (PQCWC)，介紹如何在基於雜湊密碼學的特性上做到金鑰擴展。第四節提出本研究設計的基於雜湊蝴蝶金鑰擴展機制，介紹如何在 PQCWC 基礎上做到蝴蝶金鑰擴展，把基於雜湊毛蟲金鑰擴展為基於雜湊繭公鑰，再把基於雜湊繭公鑰擴展為基於雜湊蝴蝶公鑰。第五節做性能比較，把本研究提出的 PQCWC 和基於雜湊蝴蝶金鑰擴

展機制建構在不同雜湊演算法上,並且比較計算效率。最後,第六節總結本研究的發現,並且討論未來可行的研究方向。

## 二、文獻探討

本節將先介紹 Winternitz 一次性簽章方法,再說明已知憑證請求和回應流程,以及討論其匿名性和隱私問題。最後,再介紹 IEEE 1609.2.1 設計的蝴蝶金鑰擴展機制及其在隱私保護的作法。

### 2.1、Winternitz 一次性簽章方法

Winternitz 一次性簽章方法是一個經典的基於雜湊密碼學方法,首先假設待簽署資料 $D$ 的長度限定為 $L_D$ bits,然後把每 $w_1$ bits 切成一個資料元素(element),可以把待簽署資料分為一個包含有 $m$ 個資料元素的序列,即 $D = d_1 || d_2 || ... || d_m$ 且 $m = \left\lceil \frac{L_D}{w_1} \right\rceil$。後續可以對每個資料元素各別產生其簽章值元素[22]-[23],該簽章值元素的長度 $l_s$ 根據採用的雜湊演算法而異,即總簽章長度 $L_S = m \times l_s$。詳細步驟說明如下。

#### 2.1.1、產製金鑰

在產製金鑰階段,隨機產生 $m$ 個整數元素的序列 $A$ 作為私鑰,即 $A = \{a_1, a_2, ..., a_m\}$。再對私鑰中的每一個整數元素各別做 $2^{w_1} - 1$ 次的雜湊計算,得到 $m$ 個雜湊值元素的序列 $B$ 作為公鑰 $B = \{b_1, b_2, ..., b_m\} = \{f(a_1)^{2^{w_1}-1}, f(a_2)^{2^{w_1}-1}, ..., f(a_m)^{2^{w_1}-1}\}$;其中,$f(\cdot)$ 為雜湊函數,對 $a_i$ 做 $2^{w_1} - 1$ 次雜湊計算後的值為 $b_i$。

因此,當 $w_1$ 值越大,則表示產製公鑰時需要計算的雜湊次數越多,並且會呈指數級上升。然而,當 $w_1$ 值越小,則雖然雜湊計算次數較少,產製公鑰速度較快,但序列的元素個數數量 $m$ 將會隨之變大,則私鑰長度和公鑰長度都會隨之倍數成長;並且,$w_1$ 值過小時,雖然計算速度變快,但也表示可能被破解的速度也越快,所以設置合適的值也是需要考慮的議題之一。

#### 2.1.2、產製簽章

在產製簽章階段,取得待簽署資料 $D$ 並產生其序列,由於其資料元素根據 $w_1$ bits 長度切割,所以每個資料元素的值域介於 $[0, 2^{w_1} - 1]$ 區間。可以對私鑰元素 $a_i$ 做 $d_i$ 次雜湊計算得到簽章值元素 $s_i$,依此對每個私鑰元素各別做雜湊計算得可得簽章值序列 $S = \{s_1, s_2, ..., s_m\} = \{f(a_1)^{d_1}, f(a_2)^{d_2}, ..., f(a_m)^{d_m}\}$。

其計算時間與金鑰產製類似,將取決於 $w_1$ 值;當 $w_1$ 值越大,則表示產製簽章時需要計算的雜湊次數越多,並且會呈指數級上升。同理,當 $w_1$ 值越小,則 $m$ 值越大,表示簽章值元素數數量越多,所以簽章序列的長度越長。

#### 2.1.3、驗證簽章

在驗證簽章階段,取得待簽署資料 $D$、簽章值 $S$、以及公鑰 $B$。可以對簽章值元素 $s_i$

做 $2^{w_1}-1-d_i$ 次雜湊計算得到驗章值元素 $v_i$，依此對每個簽章值元素各別做雜湊計算得可得驗章值序列 $V = \{v_1, v_2, ..., v_m\} = \{f(s_1)^{2^{w_1}-1-d_1}, f(s_2)^{2^{w_1}-1-d_2}, ..., f(s_m)^{2^{w_1}-1-d_m}\}$。當每個驗章值元素和其對應的公鑰元素都相同時，則驗章通過，即 $V = \{v_1, v_2, ..., v_m\} = \{b_1, b_2, ..., b_m\} = B$。

## 2.2、已知憑證請求和回應流程

在最基本的公開金鑰基礎建設(Public Key Infrastructure)有憑證中心可以簽發憑證給終端設備。其中，終端設備將產製憑證簽署請求(Certificate Signing Request, CSR)給憑證中心，並且在憑證簽署請求中包含終端設備資訊、終端設備公鑰、以及請求的許可(permission)列表等。由憑證中心審核終端設備資格後，簽發終端設備憑證，並且在終端設備憑證包含終端設備資訊、終端設備公鑰、以及許可列表等[24]。

除此之外，為管理公開金鑰基礎建設，將設置註冊中心，由註冊中心審核終端設備資格，再產製對應的憑證簽署請求給憑證中心，讓角色間的分工更明確。註冊中心收發終端設備訊息和驗證終端設備資格，在各式各樣不同的應用時，註冊中心扮演重要的審核角色。憑證中心負責根據憑證簽署請求來產製終端設備憑證[25]。

然而，在現行的公開金鑰基礎建設，產製出來的終端設備憑證都會包含終端設備原始的公鑰。因此，一旦有其他的攻擊者收集同一個公鑰簽發的全部訊息，將可能導致使用者隱私曝露的問題。

## 2.3、IEEE 1609.2.1 蝴蝶金鑰擴展機制

由於在車聯網環境，終端設備的隱私尤其重要，需要做好隱私保護。因此，IEEE 1609.2.1 標準中提出蝴蝶金鑰擴展機制[20],[21]，運用金鑰擴展，避免憑證中存放原始公鑰，從而達到保護隱私的目標。在此研究中僅討論顯式憑證的情境，隱式憑證不在本研究討論範圍，詳細說明如下。

### 2.3.1、產製毛蟲金鑰對

由終端設備產製兩組原始的金鑰對(key pairs)，分別是作為簽章用毛蟲金鑰對和加密用毛蟲金鑰對。其中，簽章用毛蟲金鑰對包含簽章用毛蟲私鑰 $p_1$ 和簽章用毛蟲公鑰 $P_1$；簽章用毛蟲金鑰對包含加密用毛蟲私鑰 $p_2$ 和加密用毛蟲公鑰 $P_2$。並且，IEEE 1609.2.1 標準中在非對稱式密碼學主要採用橢圓曲線密碼學，假設橢圓曲線基點是 $G$，並且橢圓曲線的階為 $n$，則簽章用毛蟲公鑰 $P_1$ 是 $p_1G$、加密用毛蟲公鑰 $P_2$ 是 $p_2G$。除此之外，為建立與註冊中心端有相同的金鑰擴展函數和參數值，所以產製兩把進階加密標準(Advanced Encryption Standard, AES)金鑰，分別是 $q_1$ 和 $q_2$。終端設備可將簽章用毛蟲公鑰 $P_1$、加密用毛蟲公鑰 $P_2$、進階加密標準金鑰 $q_1$、進階加密標準金鑰 $q_2$、以及終端設備資訊和請求的許可列表封裝成 EeRaCertRequest 封包傳送給註冊中心[20]。

### 2.3.2、產製繭公鑰

註冊中心收到 EeRaCertRequest 封包後，可以運用進階加密標準金鑰 $q_1$ 和金鑰擴展函數 $g_1(\iota, q_1)$ 產製偽隨機數 $r_1$，以及運用進階加密標準金鑰 $q_2$ 和金鑰擴展函數 $g_2(\iota, q_2)$ 產

製偽隨機數 $r_2$。其中，$\iota$為註冊中心和終端設備共同所知悉的時間週期，詳細計算定義於 IEEE 1609.2.1 標準[20]。註冊中心可以運用偽隨機數$r_1$擴展簽章用毛蟲公鑰 $P_1$ 為簽章用繭公鑰$T_1 = P_1 + r_1G$，以及運用偽隨機數 $r_2$ 擴展加密用毛蟲公鑰 $P_2$ 為加密用繭公鑰$T_2 = P_2 + r_2G$。後續再把簽章用繭公鑰$T_1$、加密用繭公鑰$T_2$、以及請求的許可列表傳送給憑證中心。

### 2.3.3、產製蝴蝶公鑰

憑證中心收到請求封包後，產生一個隨機數 $r_3$，以及運用偽隨機數 $r_3$ 擴展簽章用繭公鑰$T_1$為簽章用蝴蝶公鑰$U = T_1 + r_3G$，並產製終端設備憑證，在憑證中的公鑰欄位放置蝴蝶公鑰和許可欄位放置對應的許可列表。之後再運用加密用繭公鑰$T_2$加密終端設備憑證和隨機數 $r_3$ 得到密文 $Z$，並回傳密文 $Z$ 給註冊中心。

### 2.3.4、產製繭私鑰和蝴蝶私鑰

註冊中心收到訊息後轉送給終端設備，終端設備將運用共同已知的時間週期和擴展函數來產製偽隨機數$r_1 = g_1(\iota, q_1)$和$r_2 = g_2(\iota, q_2)$，並根據偽隨機數產製簽章用繭私鑰 $t_1 = p_1 + r_1$和加密用繭私鑰$t_2 = p_2 + r_2$。可用加密用繭私鑰$t_2$解密密文 $Z$，得到明文內容包含終端設備憑證和隨機數 $r_3$。最後再運用隨機數 $r_3$ 擴展簽章用繭私鑰$t_1$得到蝴蝶私鑰$u = t_1 + r_3$，並且蝴蝶私鑰 $u$ 和終端設備中的蝴蝶公鑰 $U$ 可以成對。在應用上，可以終端設備可以用蝴蝶私鑰 $u$ 簽署安全協定資料單元(Secure Protocol Data Unit, SPDU)，其他終端設備收到安全協定資料單元後可以用蝴蝶公鑰 $U$ 驗簽章。

## 三、本研究提出的後量子密碼學匿名憑證方案 PQCWC

有鑑於傳統憑證[24], [25]將存放終端設備原始公鑰，造成曝露隱私的問題。本研究提出 Post-Quantum Cryptography Winternitz-Chen (PQCWC)的匿名憑證方案，通過金鑰擴展機制來達到憑證中存放擴展後公鑰，而不是原始公鑰，來達到保護隱私的需求。

考慮到使用上不同的需求，本研究提出兩種 PQCWC 模型，可以供應用場域選擇合適的模型。3.1 節和 3.2 節分別說明 PQCWC 匿名憑證方案模型 1 和模型 2 流程，3.3 節從數學原理證明本研究提出的 PQCWC 匿名憑證。

本研究提出的 PQCWC 匿名憑證方案建構在下列的假設基礎上：
- 終端設備和憑證中心之間的通訊建立在安全連線通道。
- 終端設備和憑證中心具有共同已知參數$w_2$和偽隨機數產生器。
- 憑證中心有加密用金鑰對(包含私鑰 $k_{CA}$ 和公鑰 $K_{CA}$)，憑證中心有加密用金鑰對可以採用模晶格基礎金鑰封裝演算法[14]實現，以達到量子安全等級。並且，終端設備已知憑證中心加密用公鑰 $K_{CA}$。

### 3.1、PQCWC 匿名憑證方案模型 1 流程

本研究提出的 PQCWC 匿名憑證方案模型 1 流程如圖一所示，將包含終端設備產製簽章用金鑰對、憑證中心產製擴展後公鑰和終端設備匿名憑證、終端設備產製擴展後私鑰。

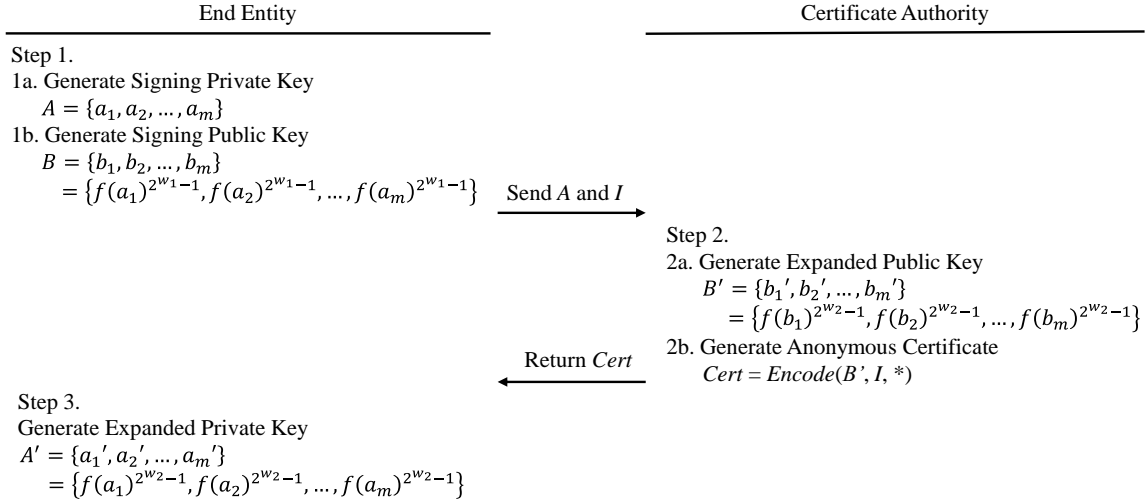

圖一: PQCWC 匿名憑證方案模型 1

### 3.1.1、終端設備產製簽章用金鑰對

由終端設備產製簽章用基於雜湊金鑰對，包含簽章用私鑰$A = \{a_1, a_2, ..., a_m\}$和簽章用公鑰$B = \{b_1, b_2, ..., b_m\} = \{f(a_1)^{2^{w_1}-1}, f(a_2)^{2^{w_1}-1}, ..., f(a_m)^{2^{w_1}-1}\}$，如 2.1 節所定義。之後終端設備再把簽章用公鑰$B$和終端設備相關待簽署資訊$I$等內容封裝成憑證請求封包發送給憑證中心。

### 3.1.2、憑證中心產製擴展後公鑰和終端設備匿名憑證

憑證中心收到憑證請求封包，確認終端設備資格無誤後，可以運用憑證中心和終端設備共同已知參數$w_2$，對簽章用公鑰$B$進行擴展，產製擴展後公鑰$B' = \{b_1', b_2', ..., b_m'\} = \{f(b_1)^{2^{w_2}-1}, f(b_2)^{2^{w_2}-1}, ..., f(b_m)^{2^{w_2}-1}\}$；其中，$f(\cdot)$為雜湊函數，對$b_i$做$2^{w_2}-1$次雜湊計算後的值為$b_i'$。憑證中心再根據擴展後公鑰$B'$和終端設備相關待簽署資訊$I$產製終端設備匿名憑證，並且回傳給終端設備。

### 3.1.3、終端設備產製擴展後私鑰

終端設備收到終端設備匿名憑證後，產製擴展後私鑰$A' = \{a_1', a_2', ..., a_m'\} = \{f(a_1)^{2^{w_2}-1}, f(a_2)^{2^{w_2}-1}, ..., f(a_m)^{2^{w_2}-1}\}$。後續終端設備可以用擴展後私鑰$A'$產製簽章，並且其他終端設備可以用擴展後公鑰$B'$來驗證簽章。

### 3.2、PQCWC 匿名憑證方案模型 2 流程

本研究提出的 PQCWC 匿名憑證方案模型 2 流程如圖二所示，將包含終端設備產製簽章用金鑰對和 AES 金鑰、憑證中心產製擴展後公鑰和終端設備匿名憑證、終端設備產製擴展後私鑰。

### 3.2.1、終端設備產製簽章用金鑰對和 AES 金鑰

由終端設備產製簽章用基於雜湊金鑰對，包含簽章用私鑰$A = \{a_1, a_2, ..., a_m\}$和簽章用公鑰$B = \{b_1, b_2, ..., b_m\} = \{f(a_1)^{2^{w_1}-1}, f(a_2)^{2^{w_1}-1}, ..., f(a_m)^{2^{w_1}-1}\}$，如 2.1 節所定義。

並且,產製一把進階加密標準金鑰 $q_3$,運用憑證中心加密用公鑰 $K_{CA}$ 對進階加密標準金鑰 $q_3$ 加密為密文 $q_3$'。之後終端設備再把簽章用公鑰 $B$、密文 $q_3$'、終端設備相關待簽署資訊 $I$ 等內容封裝成憑證請求封包發送給憑證中心。

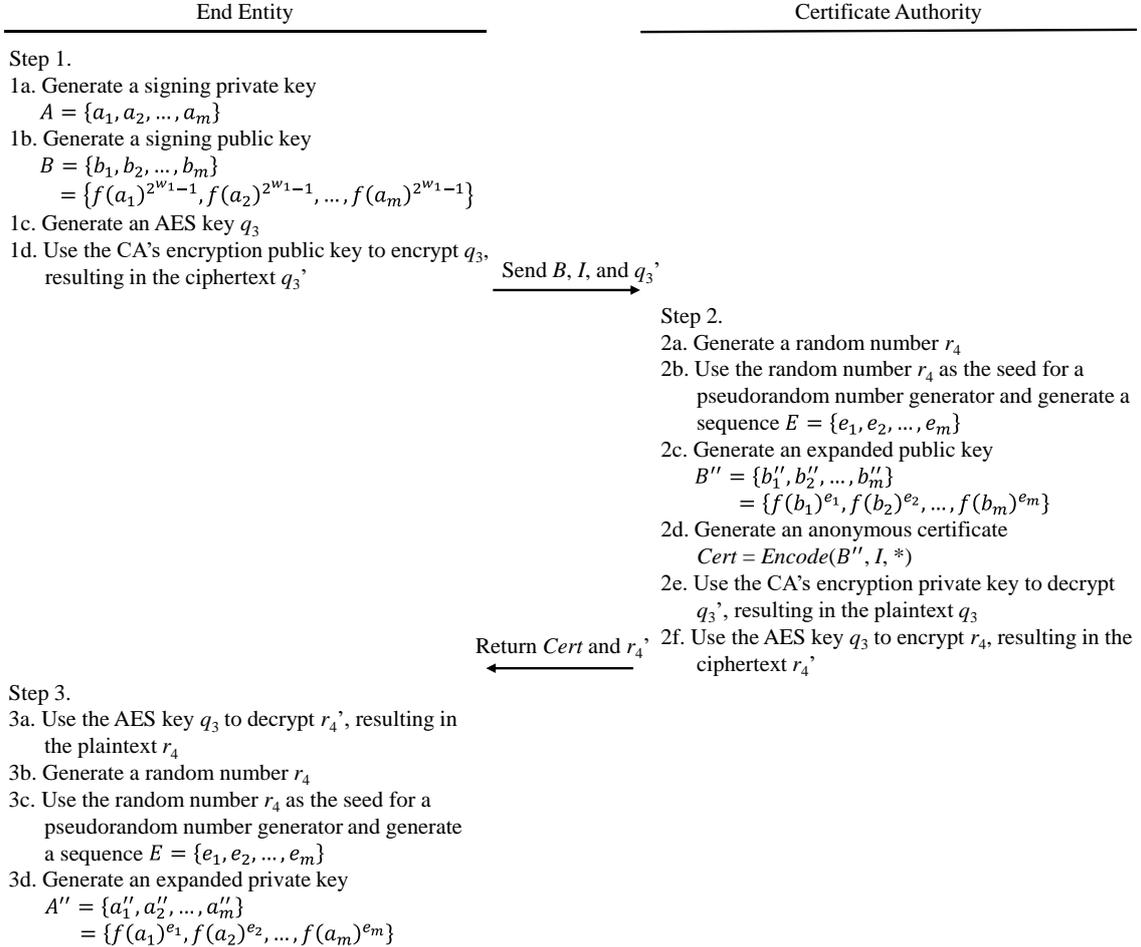

圖二: PQCWC 匿名憑證方案模型 2

### 3.2.2、憑證中心產製擴展後公鑰和終端設備匿名憑證

憑證中心收到憑證請求封包,確認終端設備資格無誤後,可以運用憑證中心加密用私鑰 $k_{CA}$ 對密文 $q_3$' 解密得到明文(即進階加密標準金鑰 $q_3$)。憑證中心可產製隨機數 $r_4$,把 $r_4$ 設定為偽隨機數產生器的隨機數種子,並產生 $m$ 個偽隨機數元素的序列 $E = \{e_1, e_2, ..., e_m\}$,並且每個偽隨機數元素的值域限定介於 $[0, 2^{w_2} - 1]$ 區間。對簽章用公鑰 $B$ 進行擴展,產製擴展後公鑰 $B'' = \{b_1'', b_2'', ..., b_m''\} = \{f(b_1)^{e_1}, f(b_2)^{e_2}, ..., f(b_m)^{e_m}\}$;其中,$f(\cdot)$ 為雜湊函數,對 $b_i$ 做 $e_i$ 次雜湊計算後的值為 $b_i''$。憑證中心再根據擴展後公鑰 $B''$ 和終端設備相關待簽署資訊 $I$ 產製終端設備匿名憑證,並且用進階加密標準金鑰 $q_3$ 加密隨機數 $r_4$ 回傳給終端設備。

### 3.2.3、終端設備產製擴展後私鑰

終端設備收到隨機數 $r_4$ 密文後,用進階加密標準金鑰 $q_3$ 解密,把 $r_4$ 設定為偽隨機

數產生器的隨機數種子，並產生 $m$ 個偽隨機數元素的序列 $E = \{e_1, e_2, ..., e_m\}$，並且每個偽隨機數元素的值域限定介於 $[0, 2^{w_2} - 1]$ 區間。對簽章用私鑰 $B$ 進行擴展，產製擴展後私鑰 $A'' = \{a_1'', a_2'', ..., a_m''\} = \{f(a_1)^{e_1}, f(a_2)^{e_2}, ..., f(a_m)^{e_m}\}$。後續終端設備可以用擴展後私鑰 $A''$ 產製簽章，並且其他終端設備可以用擴展後公鑰 $B''$ 來驗證簽章。

### 3.3、PQCWC 匿名憑證原理證明

本節將分別對提出的 PQCWC 匿名憑證方案兩個模型提供原理證明。

PQCWC 匿名憑證方案模型 1 的擴展後公鑰和擴展後私鑰的關係可以運用公式(1)證明，可以觀察到擴展後私鑰元素 $a_i'$ 做 $2^{w_1} - 1$ 次雜湊計算後為擴展後公鑰元素 $b_i'$，每個擴展後私鑰元素和擴展後公鑰元素之間分別保持 $2^{w_1} - 1$ 次雜湊計算。在產製簽章時，可以用擴展後私鑰對待簽署資料 $D = \{d_1, d_2, ..., d_m\}$ 產製簽章，如公式(2)所示。在驗證簽章時，可以用擴展後公鑰對簽章 $S = \{s_1, ..., s_m\}$ 進行驗證；如公式(3)所示，當簽章值正確時，每個驗章值元素和其對應的公鑰元素將會相同。

$$\begin{aligned} B' &= \{b_1', ..., b_m'\} \\ &= \{f(b_1)^{2^{w_2}-1}, ..., f(b_m)^{2^{w_2}-1}\} \\ &= \{f(a_1)^{2^{w_1}-1+2^{w_2}-1}, ..., f(a_m)^{2^{w_1}-1+2^{w_2}-1}\} \\ &= \{f(a_1')^{2^{w_1}-1}, ..., f(a_m')^{2^{w_1}-1}\}. \end{aligned} \quad (1)$$

$$\begin{aligned} S &= \{s_1, ..., s_m\} \\ &= \{f(a_1')^{d_1}, ..., f(a_m')^{d_m}\}. \end{aligned} \quad (2)$$

$$\begin{aligned} V &= \{v_1, ..., v_m\} \\ &= \{f(s_1)^{2^{w_1}-1-d_1}, ..., f(s_m)^{2^{w_1}-1-d_m}\} \\ &= \{f(a_1')^{2^{w_1}-1-d_1+d_1}, ..., f(a_m')^{2^{w_1}-1-d_m+d_m}\} \\ &= \{f(a_1')^{2^{w_1}-1}, ..., f(a_m')^{2^{w_1}-1}\} = B'. \end{aligned} \quad (3)$$

PQCWC 匿名憑證方案模型 2 的擴展後公鑰和擴展後私鑰的關係可以運用公式(4)證明，可以觀察到擴展後私鑰元素 $a_i''$ 做 $2^{w_1} - 1$ 次雜湊計算後為擴展後公鑰元素 $b_i''$，每個擴展後私鑰元素和擴展後公鑰元素之間分別保持 $2^{w_1} - 1$ 次雜湊計算。並且，值得一提的是 PQCWC 匿名憑證方案模型 1 可以視為 PQCWC 匿名憑證方案模型 2 的特例，當任一元素 $e_i$ 皆為 $2^{w_2} - 1$ 時的特例。在產製簽章時，可以用擴展後私鑰對待簽署資料 $D = \{d_1, d_2, ..., d_m\}$ 產製簽章，如公式(5)所示。在驗證簽章時，可以用擴展後公鑰對簽章 $S = \{s_1, ..., s_m\}$ 進行驗證；如公式(6)所示，當簽章值正確時，每個驗章值元素和其對應的公鑰元素將會相同。

$$\begin{aligned} B'' &= \{b_1'', ..., b_m''\} \\ &= \{f(b_1)^{e_1}, ..., f(b_m)^{e_m}\} \\ &= \{f(a_1)^{2^{w_1}-1+e_1}, ..., f(a_m)^{2^{w_1}-1+e_m}\} \\ &= \{f(a_1'')^{2^{w_1}-1}, ..., f(a_m'')^{2^{w_1}-1}\}. \end{aligned} \quad (4)$$

$$\begin{aligned} S &= \{s_1, ..., s_m\} \\ &= \{f(a_1'')^{d_1}, ..., f(a_m'')^{d_m}\}. \end{aligned} \quad (5)$$

$$\begin{aligned}
V &= \{v_1, \dots, v_m\} \\
&= \{f(s_1)^{2^{w_1}-1-d_1}, \dots, f(s_m)^{2^{w_1}-1-d_m}\} \\
&= \{f(a_1'')^{2^{w_1}-1-d_1+d_1}, \dots, f(a_m'')^{2^{w_1}-1-d_m+d_m}\} \\
&= \{f(a_1'')^{2^{w_1}-1}, \dots, f(a_m'')^{2^{w_1}-1}\} = B''.
\end{aligned} \quad (6)$$

## 四、本研究提出的基於雜湊蝴蝶金鑰擴展

第 3 節提出的 PQCWC 匿名憑證方案可以做到對其他終端設備匿名，但無法對憑證中心匿名。憑證中心仍可以知道擴展後公鑰和原始公鑰之間的關聯性。因此，為了做到對憑證中心匿名，本研究在 PQCWC 匿名憑證方案的基礎上提出基於雜湊蝴蝶金鑰擴展機制，通過在註冊中心做一次金鑰擴展，在憑證中心做另一次金鑰擴展，可以做到對註冊中心匿名、對憑證中心匿名、對其他終端設備匿名。以下詳細說明本研究提出的基於雜湊蝴蝶金鑰擴展機制流程和原理證明。

### 4.1、基於雜湊蝴蝶金鑰擴展機制流程

基於雜湊蝴蝶金鑰擴展機制主要建構在 PQCWC 匿名憑證方案的基礎上，並且由於 PQCWC 匿名憑證方案模型 1 是 PQCWC 匿名憑證方案模型 2 的特例，所以本節以 PQCWC 匿名憑證方案模型 2 為例說明。

本研究提出的基於雜湊蝴蝶金鑰擴展機制建構在下列的假設基礎上：
- 終端設備、註冊中心、以及憑證中心之間的通訊建立在安全連線通道。
- 終端設備和註冊中心具有共同已知時間週期$\iota$。
- 終端設備、註冊中心、以及憑證中心具有共同已知偽隨機數產生器。
- 註冊中心有加密用金鑰對，並且終端設備已知註冊中心加密用公鑰。
- 憑證中心有加密用金鑰對，並且終端設備已知憑證中心加密用公鑰。

本研究提出的基於雜湊蝴蝶金鑰擴展機制流程如圖三所示，將包含終端設備產製毛蟲金鑰對和 AES 金鑰、註冊中心產製繭公鑰、憑證中心產製蝴蝶公鑰和終端設備匿名憑證、終端設備產製蝴蝶私鑰。

#### 1) 終端設備產製毛蟲金鑰對和 AES 金鑰

由終端設備產製簽章用基於雜湊金鑰對，包含毛蟲私鑰$A = \{a_1, a_2, \dots, a_m\}$和毛蟲公鑰$B = \{b_1, b_2, \dots, b_m\} = \{f(a_1)^{2^{w_1}-1}, f(a_2)^{2^{w_1}-1}, \dots, f(a_m)^{2^{w_1}-1}\}$，如 2.1 節所定義。並且，產製兩把進階加密標準金鑰 $q_{RA}$ 和 $q_{CA}$，分別運用註冊中心加密用公鑰對進階加密標準金鑰 $q_{RA}$ 加密為密文 $q_{RA}$'，以及運用憑證中心加密用公鑰對進階加密標準金鑰 $q_{CA}$ 加密為密文 $q_{CA}$'。之後終端設備再把毛蟲公鑰$B$、密文 $q_{RA}$'、密文 $q_{CA}$'、終端設備相關待簽署資訊$I$等內容封裝成憑證請求封包發送給註冊中心。

#### 2) 註冊中心產製繭公鑰

註冊中心收到憑證請求封包，確認終端設備資格無誤後，從終端設備相關待簽署資訊$I$裡面取得待簽署請求許可列表等相關資訊$J$。註冊中心運用註冊中心加密用私鑰對密文 $q_{RA}$'解密得到明文(即進階加密標準金鑰 $q_{RA}$)。運用進階加密標準金鑰 $q_{RA}$ 對共同

已知時間週期$l$加密得到密文$r_{RA}$，把$r_{RA}$設定為偽隨機數產生器的隨機數種子，並產生$m$個偽隨機數元素的序列$E_{RA} = \{e_{RA,1}, e_{RA,2}, ..., e_{RA,m}\}$，並且每個偽隨機數元素的值域限定介於$[0, 2^{w_2} - 1]$區間。對毛蟲公鑰$B$進行擴展，產製繭公鑰$B' = \{b_1', b_2', ..., b_m'\} = \{f(b_1)^{e_{RA,1}}, f(b_2)^{e_{RA,2}}, ..., f(b_m)^{e_{RA,m}}\}$；其中，$f(\cdot)$為雜湊函數，對$b_i$做$e_{RA,i}$次雜湊計算後的值為$b_i'$。註冊中心再把繭公鑰$B'$、待簽署請求許可列表等相關資訊$J$、密文$q_{CA}'$傳送給憑證中心。

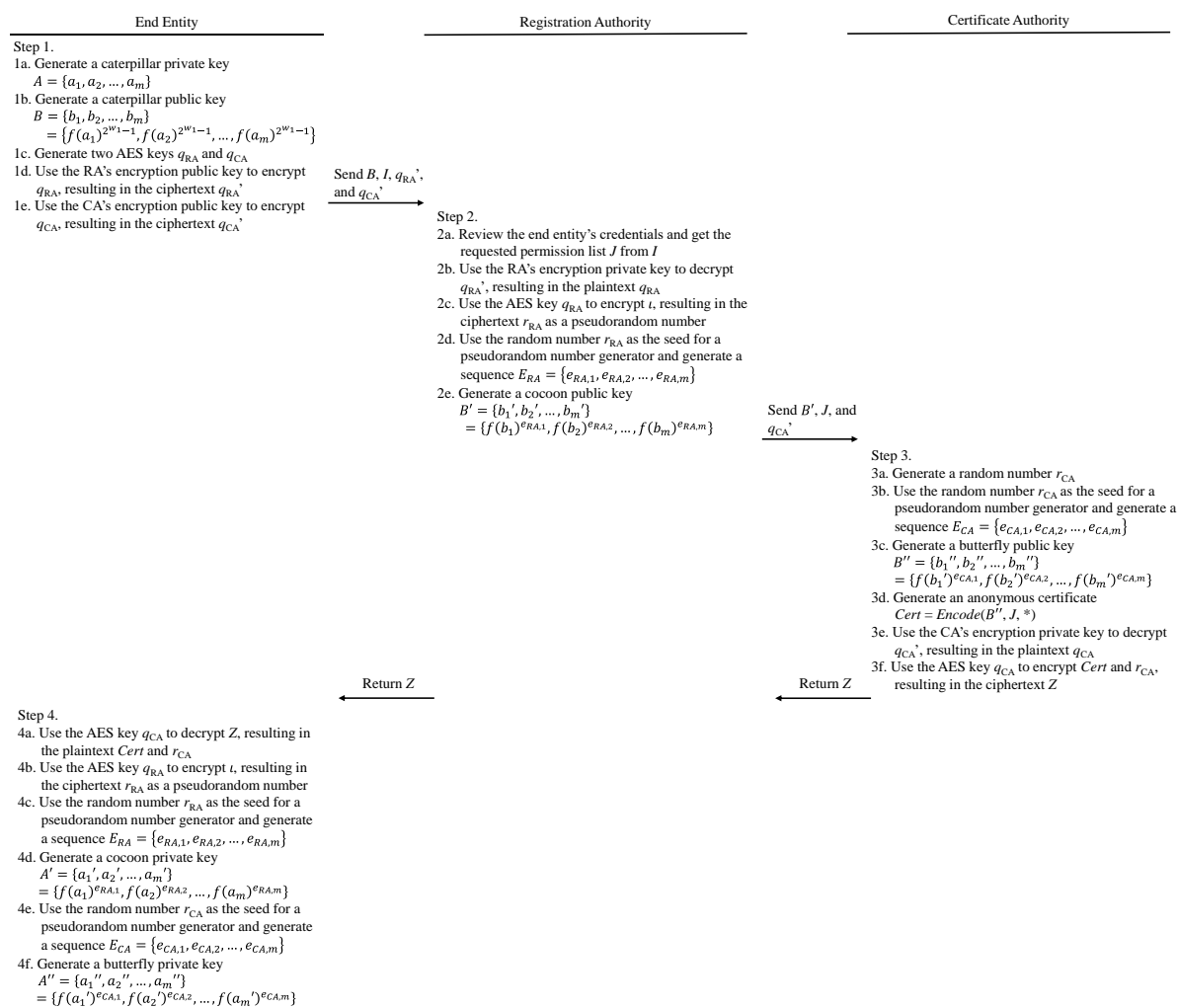

圖三：本研究提出的基於雜湊蝴蝶金鑰擴展

### 3) 憑證中心產製蝴蝶公鑰和終端設備匿名憑證

憑證中心收到封包，可以運用憑證中心加密用私鑰對密文$q_{CA}'$解密得到明文(即進階加密標準金鑰$q_{CA}$)。憑證中心可產製隨機數$r_{CA}$，把$r_{CA}$設定為偽隨機數產生器的隨機數種子，並產生$m$個偽隨機數元素的序列$E_{CA} = \{e_{CA,1}, e_{CA,2}, ..., e_{CA,m}\}$，並且每個偽隨機數元素的值域限定介於$[0, 2^{w_2} - 1]$區間。對繭公鑰$B'$進行擴展，產製蝴蝶公鑰$B'' = \{b_1'', b_2'', ..., b_m''\} = \{f(b_1')^{e_{CA,1}}, f(b_2')^{e_{CA,2}}, ..., f(b_m')^{e_{CA,m}}\}$；其中，$f(\cdot)$為雜湊函數，對$b_i'$做$e_{CA,i}$次雜湊計算後的值為$b_i''$。憑證中心再根據蝴蝶公鑰$B''$和待簽署請求許可列表等

相關資訊 $J$ 產製終端設備匿名憑證 $Cert$，並且用進階加密標準金鑰 $q_{CA}$ 加密終端設備匿名憑證 $Cert$ 和隨機數 $r_{CA}$ 為密文 $Z$ 回傳給註冊中心。

### 4) 終端設備產製蝴蝶私鑰

終端設備收到密文 $Z$ 後，用進階加密標準金鑰 $q_{CA}$ 解密取得終端設備匿名憑證 $Cert$ 和隨機數 $r_{CA}$。運用進階加密標準金鑰 $q_{RA}$ 對共同已知時間週期 $l$ 加密得到密文 $r_{RA}$，把 $r_{RA}$ 設定為偽隨機數產生器的隨機數種子，並產生 $m$ 個偽隨機數元素的序列 $E_{RA} = \{e_{RA,1}, e_{RA,2}, ..., e_{RA,m}\}$，並且每個偽隨機數元素的值域限定介於 $[0, 2^{w_2} - 1]$ 區間。對毛蟲私鑰 $A$ 進行擴展，產製繭私鑰 $A' = \{a_1', a_2', ..., a_m'\} = \{f(a_1)^{e_{RA,1}}, f(a_2)^{e_{RA,2}}, ..., f(a_m)^{e_{RA,m}}\}$；其中，$f(\cdot)$ 為雜湊函數，對 $a_i$ 做 $e_{RA,i}$ 次雜湊計算後的值為 $a_i'$。之後，把 $r_{CA}$ 設定為偽隨機數產生器的隨機數種子，並產生 $m$ 個偽隨機數元素的序列 $E_{CA} = \{e_{CA,1}, e_{CA,2}, ..., e_{CA,m}\}$，並且每個偽隨機數元素的值域限定介於 $[0, 2^{w_2} - 1]$ 區間。對繭私鑰 $A'$ 進行擴展，產製蝴蝶私鑰 $A'' = \{a_1'', a_2'', ..., a_m''\} = \{f(a_1')^{e_{CA,1}}, f(a_2')^{e_{CA,2}}, ..., f(a_m')^{e_{CA,m}}\}$；其中，$f(\cdot)$ 為雜湊函數，對 $a_i'$ 做 $e_{CA,i}$ 次雜湊計算後的值為 $a_i''$。後續終端設備可以用蝴蝶私鑰 $A''$ 產製簽章，並且其他終端設備可以用蝴蝶公鑰 $B''$ 來驗證簽章。

**4.2、基於雜湊蝴蝶金鑰擴展原理證明**

基於雜湊蝴蝶金鑰擴展的蝴蝶公鑰和蝴蝶私鑰的關係可以運用公式(7)證明，可以觀察到蝴蝶私鑰元素 $a_i''$ 做 $2^{w_1} - 1$ 次雜湊計算後為蝴蝶公鑰元素 $b_i''$，每個蝴蝶私鑰元素和蝴蝶公鑰元素之間分別保持 $2^{w_1} - 1$ 次雜湊計算。在產製簽章時，可以用蝴蝶私鑰對待簽署資料 $D = \{d_1, d_2, ..., d_m\}$ 產製簽章，如公式(8)所示。在驗證簽章時，可以用蝴蝶公鑰對簽章 $S = \{s_1, ..., s_m\}$ 進行驗證；如公式(9)所示，當簽章值正確時，每個驗章值元素和其對應的公鑰元素將會相同。

$$\begin{aligned} B'' &= \{b_1'', ..., b_m''\} \\ &= \{f(b_1')^{e_{CA,1}}, ..., f(b_m')^{e_{CA,m}}\} \\ &= \{f(b_1)^{e_{CA,1}+e_{RA,1}}, ..., f(b_m)^{e_{CA,m}+e_{RA,m}}\} \\ &= \{f(a_1)^{2^{w_1}-1+e_{CA,1}+e_{RA,1}}, ..., f(a_m)^{2^{w_1}-1+e_{CA,m}+e_{RA,m}}\} \\ &= \{f(a_1')^{2^{w_1}-1+e_{CA,1}}, ..., f(a_m')^{2^{w_1}-1+e_{CA,m}}\} \\ &= \{f(a_1'')^{2^{w_1}-1}, ..., f(a_m'')^{2^{w_1}-1}\}. \end{aligned} \quad (7)$$

$$\begin{aligned} S &= \{s_1, ..., s_m\} \\ &= \{f(a_1'')^{d_1}, ..., f(a_m'')^{d_m}\}. \end{aligned} \quad (8)$$

$$\begin{aligned} V &= \{v_1, ..., v_m\} \\ &= \{f(s_1)^{2^{w_1}-1-d_1}, ..., f(s_m)^{2^{w_1}-1-d_m}\} \\ &= \{f(a_1'')^{2^{w_1}-1-d_1+d_1}, ..., f(a_m'')^{2^{w_1}-1-d_m+d_m}\} \\ &= \{f(a_1'')^{2^{w_1}-1}, ..., f(a_m'')^{2^{w_1}-1}\} = B''. \end{aligned} \quad (9)$$

**五、實驗結果與討論**

本節將對提出的方法進行安全性比較和計算效率比較，以驗證本研究方法的可行

性。首先將說明實驗環境，之後與技術現狀(state-of-the-art; SOTA)方法進行安全性比較，以及驗證本研究方法在各種雜湊演算法的計算效率。

**5.1、實驗環境**

　　為實際驗證本研究提出後量子密碼學匿名憑證方法之效率，本研究採用一台 Windows 10 企業版的電腦模擬終端設備、註冊中心、以及憑證中心，驗證產製金鑰時間、擴展金鑰時間、產製簽章時間、驗證簽章時間。其中，實驗使用的軟硬體詳細規格是 CPU Intel(R) Core(TM) i7-10510U、記憶體 8 GB、OpenJDK 18.0.2.1、以及函式庫 Bouncy Castle Release 1.72。本研究比較各種不同的雜湊演算法，包含安全雜湊演算法 1(Secure Hash Algorithm-1, SHA-1)、SHA-2 系列、SHA-3 系列、以及 BLAKE 系列，證明提出的匿名憑證方案可以在不增加金鑰長度、簽章長度、產製金鑰時間、產製簽章時間、驗證簽章時間的情況下達到匿名性。

**5.2、安全性比較**

　　在安全性的部分，由於 IEEE 1609.2.1 標準中的蝴蝶金鑰擴展機制可以做到匿名性，包含可以對其他終端設備匿名和對憑證中心匿名[20], [21]。然而，IEEE 1609.2.1 標準中的蝴蝶金鑰擴展機制主要建構在橢圓曲線密碼學的基礎上，所以將可能受到量子計算的攻擊，無法達到量子安全等級。在[22], [23]的方案，主要建構在基於雜湊密碼學，可以達到量子安全等級，然而在憑證中主要放原始公鑰，將造成隱私威脅，比較結果如表一所示。

　　本研究提出的 PQCWC 匿名憑證方案主要建構在基於雜湊密碼學，可以達到量子安全等級，並且做到金鑰擴展，所以在未搭配蝴蝶金鑰擴展機制的情況下，也能對其他終端設備匿名。除此之外，本研究在 PQCWC 匿名憑證方案基礎上提出基於雜湊蝴蝶金鑰擴展機制，可以做到對憑證中心匿名，達到更好的隱私保護。

表一: 安全性比較結果

| 方法 | 量子安全 | 隱私保護 | |
|---|---|---|---|
| | | 對其他終端設備匿名 | 對憑證中心匿名 |
| [20] | | v | v |
| [21] | | v | v |
| [22] | v | | |
| [23] | v | | |
| 本研究提出的 PQCWC 匿名憑證方案模型 1 | v | v | |
| 本研究提出的 PQCWC 匿名憑證方案模型 2 | v | v | |
| 本研究提出的基於雜湊蝴蝶金鑰擴展 | v | v | v |

**5.3、簽章長度比較**

　　在實驗環境中，本研究假設待簽署資料 $D$ 是固定長度 256 位元(可以是原始資料經過 SHA-256 計算後的 digest 值)，並且每一個私鑰元素值 $a_i$ 長度為 256 位元，並且分別

考慮 $w_1$ 長度為 8 位元和 16 位元的情況下做比較，簽章長度如表二和表三所示。由於私鑰序列、公鑰序列、以及簽章序列內的元素個數 $m$ 主要取決於待簽署資料 $D$ 長度和 $w_1$ 長度。並且，根據雜湊演算法會決定簽章序列中的每個元素的長度，從而決定簽章序列總長度。例如，當 $w_1$ 為 8 位元時，則 $m$ 為 32，並且 SHA-1 的雜湊值長度是 160 位元，所以簽章序列總共包含 32 個 160 位元的雜湊值，所以總長度是 5120 位元(即 640 位元組)。從表二和表三可以觀察到，本研究提出的方法並不會增加簽章長度，可以在同樣長度下做到匿名性。另外，雖然隨著 $w_1$ 長度增加可以減少簽章長度，但計算時間將指數級增加，在後續章節將深入討論。

表二: 簽章長度比較結果($w_1 = 8$)(單位：位元)

| Hash Algorithm | Hash Value Length | $w_1$ | $m$ | [22], [23] | The Proposed Method |
|---|---|---|---|---|---|
| SHA-1 | 160 | 8 | 32 | 5120 | 5120 |
| SHA-224 | 224 | 8 | 32 | 7168 | 7168 |
| SHA-256 | 256 | 8 | 32 | 8192 | 8192 |
| SHA-384 | 384 | 8 | 32 | 12288 | 12288 |
| SHA-512 | 512 | 8 | 32 | 16384 | 16384 |
| SHA3-224 | 224 | 8 | 32 | 7168 | 7168 |
| SHA3-256 | 256 | 8 | 32 | 8192 | 8192 |
| SHA3-384 | 384 | 8 | 32 | 12288 | 12288 |
| SHA3-512 | 512 | 8 | 32 | 16384 | 16384 |
| BLAKE2-256 | 256 | 8 | 32 | 8192 | 8192 |
| BLAKE2-384 | 384 | 8 | 32 | 12288 | 12288 |
| BLAKE2-512 | 512 | 8 | 32 | 16384 | 16384 |

表三: 簽章長度比較結果($w_1 = 16$)(單位：位元)

| Hash Algorithm | Hash Value Length | $w_1$ | $m$ | [22], [23] | The Proposed Method |
|---|---|---|---|---|---|
| SHA-1 | 160 | 16 | 16 | 2560 | 2560 |
| SHA-224 | 224 | 16 | 16 | 3584 | 3584 |
| SHA-256 | 256 | 16 | 16 | 4096 | 4096 |
| SHA-384 | 384 | 16 | 16 | 6144 | 6144 |
| SHA-512 | 512 | 16 | 16 | 8192 | 8192 |
| SHA3-224 | 224 | 16 | 16 | 3584 | 3584 |
| SHA3-256 | 256 | 16 | 16 | 4096 | 4096 |
| SHA3-384 | 384 | 16 | 16 | 6144 | 6144 |
| SHA3-512 | 512 | 16 | 16 | 8192 | 8192 |
| BLAKE2-256 | 256 | 16 | 16 | 4096 | 4096 |
| BLAKE2-384 | 384 | 16 | 16 | 6144 | 6144 |
| BLAKE2-512 | 512 | 16 | 16 | 8192 | 8192 |

值得一提的是雖然私鑰序列長度也是 $m$ 個元素，但可以只存一個 256 位元長度的隨機數作為偽隨機數產生器的隨機數種子，後續用偽隨機數產生器產生 $m$ 個元素作為私鑰。而公鑰序列雖然也是 $m$ 個元素，但可以對整個公鑰序列做一次 SHA-256 計算，就可以把公鑰長度降為 256 位元。因此，在基於雜湊密碼學的私鑰和公鑰都可以較簡短，所以本研究不對私鑰和公鑰長度深入討論。

### 5.4、產製金鑰與擴展金鑰比較

本節為了對比金鑰產製的計算時間和本研究提出的 PQCWC 匿名憑證方案(即金鑰擴展)的計算時間，分別採用不同的雜湊演算法(即前面章節所提的雜湊函數$f(\cdot)$)來產製 1000 把金鑰和產製 1000 把擴展金鑰，並且為了公平比較，本節採用的 $w_1$ 值和 $w_2$ 值一樣，單一元素實驗結果分別如表四和表五所示。由實驗結果可以觀察到，由於金鑰產製的計算時間主要取決於$2^{w_1}-1$次雜湊計算，而金鑰擴展的計算時間主要取決於$2^{w_2}-1$次雜湊計算。因此，當 $w_1$ 值和 $w_2$ 值一樣時，則金鑰產製時間和 PQCWC 匿名憑證方案模型 1 的金鑰擴展時間差異不大。此外，由於 PQCWC 匿名憑證方案模型 2 是採用介於$[0, 2^{w_2}-1]$區間的隨機數來擴展，所以計算時間低於金鑰產製時間。為更清晰觀察不同方案的差異，圖四和圖五呈現實驗結果的盒鬚圖。由圖四和圖五可以觀察到，金鑰產製時間和 PQCWC 匿名憑證方案模型 1 的金鑰擴展時間的資料分佈類似。然而，在 PQCWC 匿名憑證方案模型 2 金鑰擴展時間的平均值和中位數皆低於金鑰產製時間，但最大值和最小值的分佈較廣泛，原因在於其隨機數介於$[0, 2^{w_2}-1]$，影響計算時間落差較大。

表四：產製金鑰和擴展金鑰比較結果($w_1 = 8$)(單位：毫秒)

| Hash Algorithm | Key Generation | Key Expansion Based on PQCWC Scheme 1 | Key Expansion Based on PQCWC Scheme 2 |
| --- | --- | --- | --- |
| SHA-1 | 0.085 | 0.079 | 0.041 |
| SHA-224 | 0.110 | 0.104 | 0.053 |
| SHA-256 | 0.129 | 0.120 | 0.059 |
| SHA-384 | 0.166 | 0.155 | 0.077 |
| SHA-512 | 0.146 | 0.142 | 0.072 |
| SHA3-224 | 0.250 | 0.241 | 0.122 |
| SHA3-256 | 0.302 | 0.286 | 0.141 |
| SHA3-384 | 0.220 | 0.212 | 0.111 |
| SHA3-512 | 0.324 | 0.313 | 0.160 |
| BLAKE2-256 | 0.130 | 0.121 | 0.060 |
| BLAKE2-384 | 0.139 | 0.132 | 0.067 |
| BLAKE2-512 | 0.161 | 0.154 | 0.073 |

除此之外，為客觀驗證差異的顯著性，本研究採用統計 *t*-檢定，計算不同方案之間的兩兩對比結果，由於總共有 12 個演算法的計算結果，所以自由度為 22，觀察當 *t*-value

值大於 2.074 時表示有顯著差異[26]。$t$-檢定實驗結果分別如表六和表七所示。由實驗結果可以觀察到，金鑰產製時間和 PQCWC 匿名憑證方案模型 1 的金鑰擴展時間沒有顯著差異。然而，由於 PQCWC 匿名憑證方案模型 2 金鑰擴展時間和金鑰產製時間、PQCWC 匿名憑證方案模型 1 金鑰擴展時間都有顯著差異，並且由表四和表五知道 PQCWC 匿名憑證方案模型 2 金鑰擴展時間是顯著較低的。因此，本研究提出的 PQCWC 匿名憑證方案可以具備更高的計算效率。

表五：產製金鑰和擴展金鑰比較結果($w_1 = 16$)(單位：毫秒)

| Hash Algorithm | Key Generation | Key Expansion Based on PQCWC Scheme 1 | Key Expansion Based on PQCWC Scheme 2 |
|---|---|---|---|
| SHA-1 | 15.049 | 14.902 | 7.365 |
| SHA-224 | 15.764 | 15.608 | 7.727 |
| SHA-256 | 15.540 | 15.379 | 7.756 |
| SHA-384 | 18.579 | 18.470 | 8.902 |
| SHA-512 | 19.138 | 19.368 | 9.600 |
| SHA3-224 | 55.740 | 55.634 | 28.029 |
| SHA3-256 | 51.622 | 51.787 | 26.001 |
| SHA3-384 | 51.501 | 51.803 | 26.327 |
| SHA3-512 | 51.961 | 52.158 | 26.054 |
| BLAKE2-256 | 28.725 | 28.741 | 14.370 |
| BLAKE2-384 | 28.464 | 28.368 | 14.290 |
| BLAKE2-512 | 28.536 | 28.328 | 14.289 |

表六：產製金鑰和擴展金鑰 $t$-檢定結果($w_1 = 8$)

| $t$-value | Key Generation | Key Expansion Based on PQCWC Scheme 1 |
|---|---|---|
| Key Expansion Based on PQCWC Scheme 1 | 0.279 | |
| Key Expansion Based on PQCWC Scheme 2 | **3.807** | **3.545** |

表七：產製金鑰和擴展金鑰 $t$-檢定結果($w_1 = 16$)

| $t$-value | Key Generation | Key Expansion Based on PQCWC Scheme 1 |
|---|---|---|
| Key Expansion Based on PQCWC Scheme 1 | 0.001 | |
| Key Expansion Based on PQCWC Scheme 2 | **2.992** | **2.977** |

(a). 金鑰產製時間

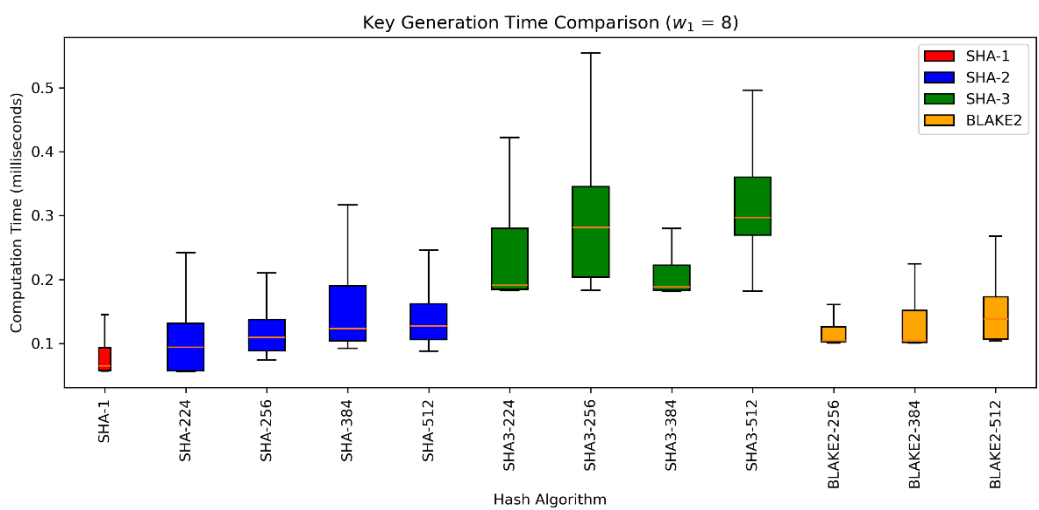

(b). PQCWC 匿名憑證方案模型 1 金鑰擴展時間

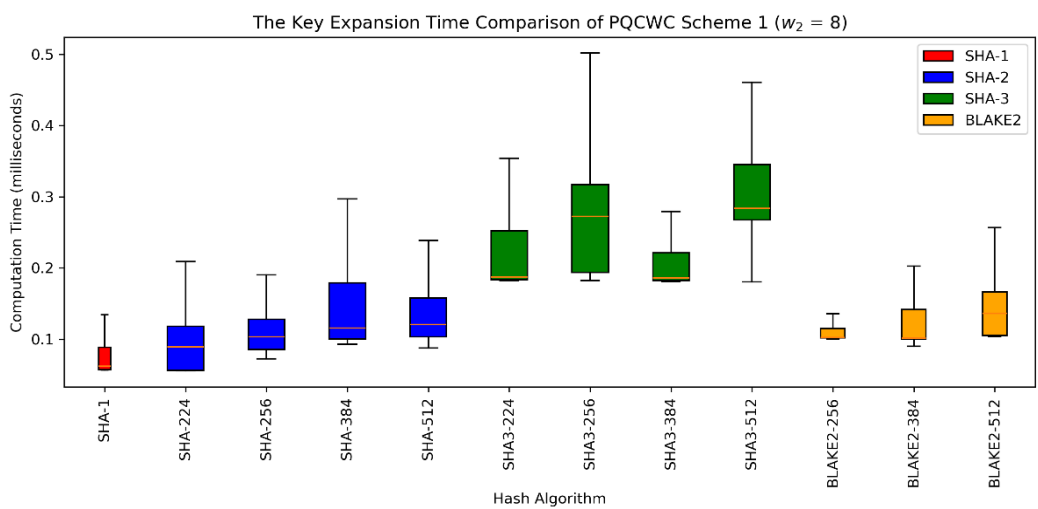

(c). PQCWC 匿名憑證方案模型 2 金鑰擴展時間

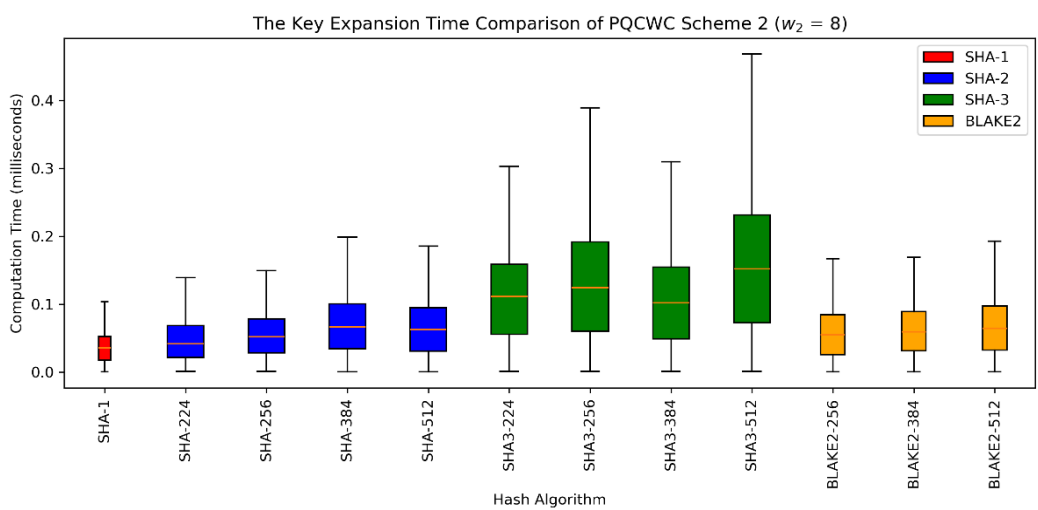

圖四: 產製金鑰和擴展金鑰比較($w_1 = 8$)(單位：毫秒).

(a). 金鑰產製時間

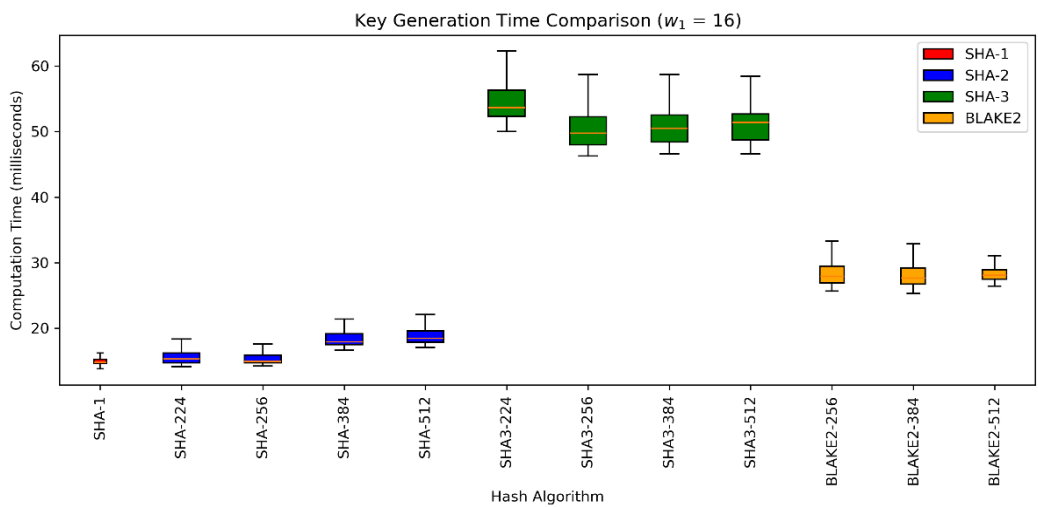

(b). PQCWC 匿名憑證方案模型 1 金鑰擴展時間

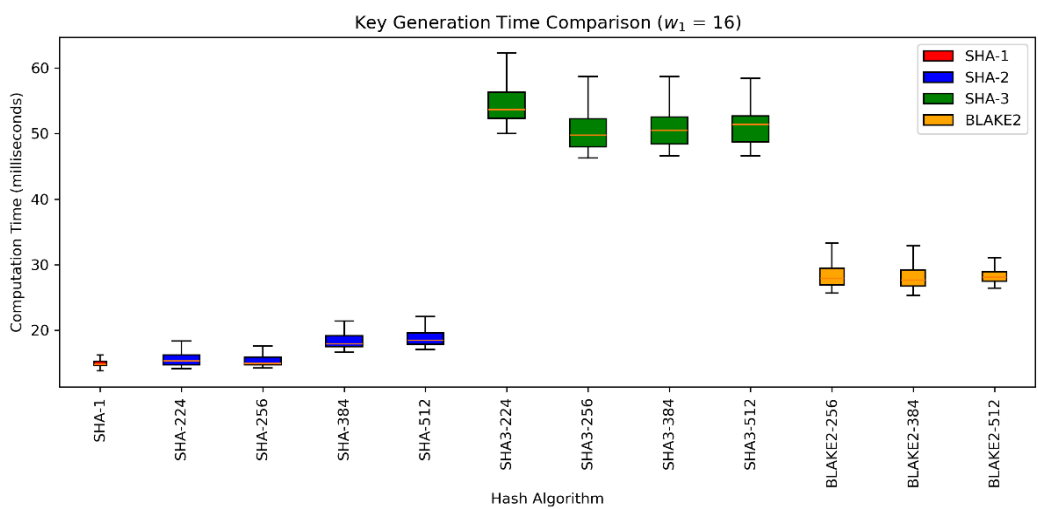

(c). PQCWC 匿名憑證方案模型 2 金鑰擴展時間

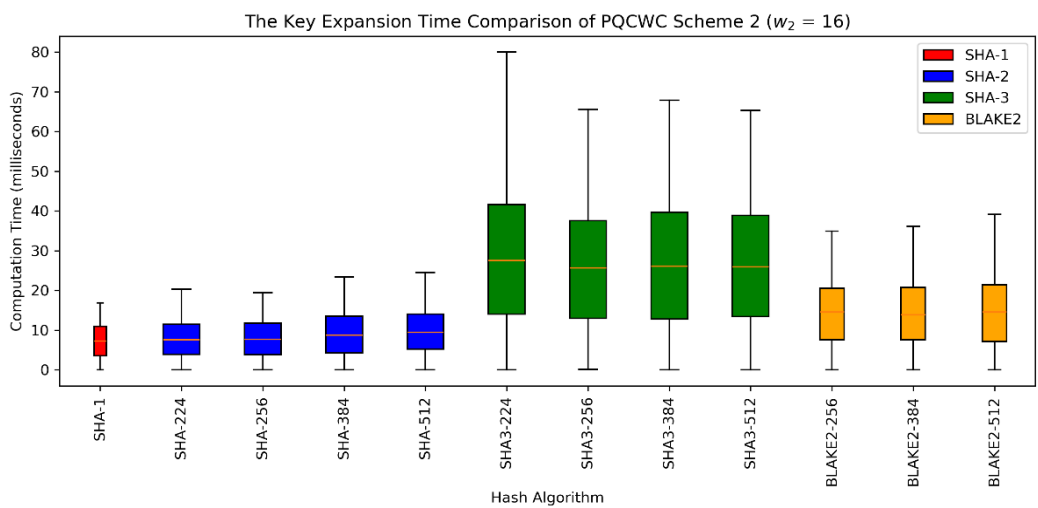

圖五: 產製金鑰和擴展金鑰比較($w_1 = 16$)(單位:毫秒).

## 5.5、產製簽章比較

本節為了對比用原始私鑰產製簽章的計算時間和用本研究提出的 PQCWC 匿名憑證方案(即擴展後私鑰)產製簽章的計算時間，分別採用不同的雜湊演算法來產製 1000 個簽章，實驗結果分別如表八和表九所示。由實驗結果可以觀察到，不論用原始私鑰或是擴展後私鑰，私鑰元素根據待簽署資料 $D$ 的元素值來決定$d_i$次雜湊計算。因此，不論採用原始私鑰、PQCWC 匿名憑證方案模型 1 擴展後私鑰、PQCWC 匿名憑證方案模型 2 擴展後私鑰的計算時間都類似。$t$-檢定實驗結果分別如表十和表十一所示，結果顯示皆無顯著差異。

表八: 產製簽章比較結果($w_1 = 8$)(單位：毫秒)

| Hash Algorithm | [22], [23] | Signature Generation Based on PQCWC Scheme 1 | Signature Generation Based on PQCWC Scheme 2 |
| --- | --- | --- | --- |
| SHA-1 | 0.041 | 0.038 | 0.041 |
| SHA-224 | 0.051 | 0.051 | 0.047 |
| SHA-256 | 0.061 | 0.062 | 0.059 |
| SHA-384 | 0.074 | 0.073 | 0.075 |
| SHA-512 | 0.070 | 0.067 | 0.065 |
| SHA3-224 | 0.128 | 0.119 | 0.119 |
| SHA3-256 | 0.149 | 0.142 | 0.140 |
| SHA3-384 | 0.108 | 0.112 | 0.109 |
| SHA3-512 | 0.155 | 0.155 | 0.156 |
| BLAKE2-256 | 0.060 | 0.058 | 0.058 |
| BLAKE2-384 | 0.068 | 0.068 | 0.067 |
| BLAKE2-512 | 0.077 | 0.075 | 0.074 |

表九: 產製簽章比較結果($w_1 = 16$)(單位：毫秒)

| Hash Algorithm | [22], [23] | Based on PQCWC Scheme 1 | Based on PQCWC Scheme 2 |
| --- | --- | --- | --- |
| SHA-1 | 7.580 | 7.576 | 7.520 |
| SHA-224 | 7.832 | 7.872 | 7.851 |
| SHA-256 | 7.499 | 7.502 | 7.484 |
| SHA-384 | 9.234 | 9.217 | 9.182 |
| SHA-512 | 9.504 | 9.459 | 9.482 |
| SHA3-224 | 29.137 | 29.099 | 29.138 |
| SHA3-256 | 25.430 | 25.474 | 25.420 |
| SHA3-384 | 25.828 | 25.787 | 25.721 |
| SHA3-512 | 26.247 | 26.149 | 26.110 |
| BLAKE2-256 | 14.407 | 14.451 | 14.393 |
| BLAKE2-384 | 14.393 | 14.408 | 14.381 |
| BLAKE2-512 | 14.622 | 14.633 | 14.600 |

表十: 產製簽章 $t$-檢定結果($w_1 = 8$)

| t-value | [22], [23] | Based on PQCWC Scheme 1 |
|---|---|---|
| Based on PQCWC Scheme 1 | 0.119 | |
| Based on PQCWC Scheme 2 | 0.182 | 0.064 |

表十一: 產製簽章 $t$-檢定結果($w_1 = 16$)

| t-value | [22], [23] | Based on PQCWC Scheme 1 |
|---|---|---|
| Based on PQCWC Scheme 1 | 0.002 | |
| Based on PQCWC Scheme 2 | 0.011 | 0.008 |

### 5.6、驗證簽章比較

本節為了對比用原始公鑰驗證簽章的計算時間和用本研究提出的 PQCWC 匿名憑證方案(即擴展後公鑰)驗證簽章的計算時間，分別採用不同的雜湊演算法來驗證前一節產製的 1000 個簽章，實驗結果分別如表十二和表十三所示。由實驗結果可以觀察到，不論用原始公鑰或是擴展後公鑰，簽章元素皆根據待簽署資料 $D$ 的元素值來決定 $2^{w_1} - 1 - d_i$ 次雜湊計算。因此，不論採用原始公鑰、PQCWC 匿名憑證方案模型 1 擴展後公鑰、PQCWC 匿名憑證方案模型 2 擴展後公鑰的計算時間都類似。$t$-檢定實驗結果分別如表十四和表十五所示，結果顯示皆無顯著差異。

表十二: 驗證簽章比較結果($w_1 = 8$)(單位：毫秒)

| Hash Algorithm | [22], [23] | Signature Generation Based on PQCWC Scheme 1 | Signature Generation Based on PQCWC Scheme 2 |
|---|---|---|---|
| SHA-1 | 0.040 | 0.042 | 0.039 |
| SHA-224 | 0.051 | 0.051 | 0.050 |
| SHA-256 | 0.060 | 0.057 | 0.057 |
| SHA-384 | 0.077 | 0.076 | 0.073 |
| SHA-512 | 0.070 | 0.069 | 0.068 |
| SHA3-224 | 0.118 | 0.120 | 0.118 |
| SHA3-256 | 0.145 | 0.137 | 0.141 |
| SHA3-384 | 0.105 | 0.108 | 0.104 |
| SHA3-512 | 0.160 | 0.157 | 0.155 |
| BLAKE2-256 | 0.061 | 0.060 | 0.060 |
| BLAKE2-384 | 0.067 | 0.068 | 0.066 |
| BLAKE2-512 | 0.080 | 0.077 | 0.077 |

表十三: 驗證簽章比較結果($w_1 = 16$)(單位：毫秒)

| Hash Algorithm | [22], [23] | Based on PQCWC Scheme 1 | Based on PQCWC Scheme 2 |
|---|---|---|---|
| SHA-1 | 7.489 | 7.469 | 7.443 |
| SHA-224 | 7.910 | 7.940 | 7.939 |
| SHA-256 | 8.007 | 8.053 | 7.998 |
| SHA-384 | 9.419 | 9.408 | 9.453 |
| SHA-512 | 9.650 | 9.623 | 9.643 |
| SHA3-224 | 26.767 | 26.578 | 26.517 |
| SHA3-256 | 26.364 | 26.362 | 26.215 |
| SHA3-384 | 25.914 | 25.881 | 25.771 |
| SHA3-512 | 26.093 | 25.942 | 25.995 |
| BLAKE2-256 | 14.563 | 14.593 | 14.453 |
| BLAKE2-384 | 14.230 | 14.254 | 14.102 |
| BLAKE2-512 | 14.120 | 13.954 | 13.913 |

表十四: 驗證簽章 $t$-檢定結果($w_1 = 8$)

| $t$-value | [22], [23] | Based on PQCWC Scheme 1 |
|---|---|---|
| Based on PQCWC Scheme 1 | 0.067 | |
| Based on PQCWC Scheme 2 | 0.146 | 0.081 |

表十五: 驗證簽章 $t$-檢定結果($w_1 = 16$)

| $t$-value | [22], [23] | Based on PQCWC Scheme 1 |
|---|---|---|
| Based on PQCWC Scheme 1 | 0.012 | |
| Based on PQCWC Scheme 2 | 0.027 | 0.016 |

## 六、結論與未來研究

　　本研究提出後量子密碼學匿名憑證方案 PQCWC，可以達到量子安全等級，同時又可以提供匿名性保護隱私。除此之外，本研究在後量子密碼學匿名憑證方案 PQCWC 基礎上再提出基於雜湊蝴蝶金鑰擴展機制，進一步做到對憑證中心匿名，充份保護終端設備的隱私。在實驗章節，本研究證明提出的方法與現有技術在簽章長度、金鑰產製時間和金鑰擴展時間、簽章產製時間、以及簽章驗證時間都沒有顯著差異。可以在相同計算時間的情況下，提供匿名性。在未來研究可以考量延伸本研究提出的後量子密碼學匿名憑證方案 PQCWC 加入 Merkle 樹，以及結合 SLH-DSA 演算法[12]來提升實用性。

## 參考文獻